\documentclass[pre,aps,twocolumn,amsmath,floatfix]{revtex4}

\usepackage{graphicx}

\begin{document}

\newcommand \be {\begin{equation}}
\newcommand \ee {\end{equation}}
\newcommand \bea {\begin{eqnarray}}
\newcommand \eea {\end{eqnarray}}

\title{Dynamic phase diagram of the Number Partitioning Problem}
\author{Ivan Junier$^1$ and Eric Bertin$^2$}
\affiliation{
$^1$PMMH, ESPCI, 10 rue Vauquelin, F-75005 Paris, France\\
$^2$Department of Theoretical Physics, University of Geneva,
CH-1211 Geneva 4, Switzerland}
\date{\today}

\begin{abstract}
We study the dynamic phase diagram of a spin model associated with the Number Partitioning Problem,
as a function of temperature and of the fraction $K/N$ of spins allowed to flip simultaneously.
The case $K=1$ reproduces the activated behavior of Bouchaud's trap model, whereas the opposite limit $K=N$ can be mapped onto the entropic trap model proposed by Barrat and M\'ezard.
In the intermediate case $1 \ll K \ll N$, the dynamics corresponds to a modified version of the Barrat and M\'ezard model, which includes a slow (rather than instantaneous) decorrelation at each step. A transition from an activated regime to an entropic one is observed at temperature $T_g/2$ in agreement 
with recent work on this model.
Ergodicity breaking occurs for $T<T_g/2$ in the thermodynamic limit, if $K/N \to 0$.
In this temperature range, the 
model exhibits a non trivial fluctuation-dissipation relation leading for $K \ll N$
to a single effective temperature equal to $T_g/2$.
These results give new insights on the relevance and limitations of the picture proposed by simple trap models.

\end{abstract}
\maketitle

\section{Introduction}

An important step towards the understanding of glassy dynamics \cite{Review} has
been made when it was recognized that some generic properties of
configuration space --or phase space-- could be responsible for
the dramatic slowing down of the dynamics \cite{Stil95,Heuer99,Deben01,Scior02}. In
particular, the geometric structure of phase space leads
schematically to two different kinds of dynamics: an `activated'
dynamics in which the system is trapped in local minima by significant energy
barriers, and an `entropic' dynamics which results
from a decreasing number of downwards directions when visiting saddles in
configuration space \cite{lalou,Scior00,Broderix,Giardina,Keyes,Angelani03}. In this latter case, the system spends
most of its time wandering in search of these rare paths which
would allow it to decrease its energy.

A popular and qualitative
description of these glassy behaviors has been proposed in the past decade in terms of trap
models, in which a very simplified phase space dynamics takes place.
In these models, any state can be reached from any
other through a single transition, disregarding any non trivial structure related to the finite dimensionality of real space.
Such models actually focus on
the distribution of low energy states, often assumed to be
exponential, which can be justified on the basis
of extreme statistics \cite{BouchMez}.

Depending on the specific choice of the
transition rates, one can build an activated dynamics --as in
Bouchaud's trap model (BTM) \cite{Bouchaud,BD,Monthus}-- or an entropic
one --as in the Barrat and M\'ezard model (BMM) \cite{BMM,Bertin}.
Considering a finite size BMM, or introducing by hand a threshold level,
one can observe a crossover from an entropic to an activated regime \cite{Bertin}.
Intuitively, such a crossover means that the system is no longer able to find downwards directions since it has reached the bottom of the `valley'.
Further evolution can only proceed by crossing energy barriers.

In spite of the conceptual interest of these models, it seems
rather difficult to find microscopic models (i.e. models in which
microscopic degrees of freedom are explicitly described) where a
reasonably clear mapping to such trap models can be proposed. This
situation is particularly striking since the physical
interpretation of trap models looks quite clear, but
arguments usually fail to go beyond a qualitative level.

The first explicit (and mathematically rigorous) mapping \cite{BenArous} was
proposed between the {\it finite size} Random Energy Model
\cite{Derrida} and the BTM. Trap mechanism has also been shown to be
a tangible description of supercooled liquids slowing down
when considering the distribution of the energy associated to the inherent structures \cite{Reich}.
On the other hand, it has been proposed
recently \cite{NPPtrap} to use a modified version of the Number
Partitioning Problem (NPP), mapped onto a fully connected spin model
with a one-spin-flip dynamics, to illustrate how an activated
behavior typical of the BTM arises from a microscopic dynamics.

In the present paper, we discuss the influence of the choice of
the dynamics on the behavior of the NPP. We show in particular
that varying the number of spins that can be flipped
simultaneously allows to recover most of phenomenology of glass
theory, namely transitions between entropic and activated
behavior, non-linear as well as linear (with non-trivial slope)
fluctuation-dissipation relation (FDR), and ergodicity
breaking. Conversely, these microscopic realizations allow to shed
some new light on the interpretation, as well as limitations,
of simple phase space models like the BTM and the BMM.

The paper is organized as follows: Sect.~II introduces the NPP model, and Sect.~III describes the basic mappings onto usual trap models for some specific dynamical rules. In Sect.~IV, we introduce more general dynamical rules, and study the behavior of the model, emphasizing the entropic-to-activated transitions as well as relations to trap models. In Sect.~V, the FDR is studied and shown to be linear in a particular limit, with a non trivial slope. Finally, we discuss in Sect.~VI the interpretation of this linear FDR, as well as the influence of the energy density on the transition between entropic and activated regimes.

\section{Optimization and spin model} \label{sect-opt}

The optimization problem of the unconstrained NPP, described by a cost function, can be mapped onto a fully connected spin model with disordered Hamiltonian \cite{Mertens,Mertens2,Silvio}.
By suitably choosing the cost function and the dynamics, this spin model can be given a glassy behavior which resembles closely that of the BTM \cite{NPPtrap}.
Given a set of $N$ random
real numbers $a_1, a_2, ... a_N$ uniformly drawn from the interval $[0,1]$,
the original NPP consists of finding the optimum configuration $\{ s_i
\}_{i=1...N}^{opt}$, where $s_i=\pm 1$ are Ising spins, which minimizes the following cost function:

\begin{equation}
E_m=\left| \sum_{i=1}^N a_i s_i \right|
\end{equation}
This is equivalent to finding a partition of the set $\{a_i\}$ into two subsets ${\cal S}_1$ and ${\cal S}_2$ such that the sums of the $a_i$'s within each subset are as close as possible.
In terms of spin systems, this problem corresponds without loss
of generality to an Hamiltonian $E_{\rm Mattis}=E_m^2$ describing an
antiferromagnetic Mattis-like spin-glass:

\begin{equation}
E_{\rm Mattis}=\sum_{i,j} a_i a_j s_i s_j
\end{equation}

From a thermodynamic point of view, Mertens \cite{Mertens,Mertens2} has
shown that the ground state of this Hamiltonian was $\langle
E_{m_o} \rangle =\sqrt{\frac{2}{3}\pi N} 2^{-N}$ so that no glass
transition at finite temperature is expected in the thermodynamic
limit. Interestingly, from such an Hamiltonian, one can derive a
new cost function, i.e.~a new energy, that has an extensive
ground state:

\begin{equation}
E= T_g \ln (E_m) =  T_g \ln \left| \sum_{i=1}^N a_i s_i \right|
\label{Ham}
\end{equation}
where $T_g$ fixes the energy scale; from now on the ground state
scales like $\langle E_o \rangle \sim - N \, T_g \ln 2$. In
this paper, we consider the study of a system defined by
such an Hamiltonian.

In this system, energies $E < k \ln N$ (where $k$ is some positive constant) are independent random
variables \cite{Mertens,NPPtrap} that are distributed according to (assuming $N \gg 1$):
\begin{equation}
\rho(E)= \mathcal{N}\exp \left(\beta_g E  - \frac{1}{2 \sigma^2 N} \exp \left(2
\beta_g E\right)\right)
\label{energie}
\end{equation}
with $\mathcal{N}=2 \beta_g / \sqrt{2 \pi \sigma^2 N}$, and $\beta_g=T_g^{-1}$.
The essential property of this distribution $\rho(E)$ is that it has an exponential tail for $E \to -\infty$. Such a tail is usually the key ingredient to obtain a glass transition at finite temperature.

Using Derrida's microcanonical argument for the random energy model
(REM) \cite{Derrida}, a thermodynamic transition is expected at
temperature $T_g$, below which the system is frozen in a limited number of states surrounding the ground state, so that the entropy (density) vanishes.

The glass transition in the present model resembles the standard REM
transition, the only difference being that the former is first
order whereas the latter is second order \cite{BouchMez,NPPtrap}. In
particular, an important property shared by both models is that
for low energy states, magnetization and energy become
decorrelated.

From an optimization point of view, many interesting questions are
inherent to the NP-hard nature of the NPP
\cite{Mertens,micro}. Thus, it is interesting to study, when
$T<T_g$, how the system (\ref{Ham}) approaches the ground state
depending on the local dynamical laws, given the prescription of
detailed balance. Since the NPP belongs to the class of
NP-complete problems, time needed for any algorithm to get the
perfect partition is exponential in the system size $N$. The most
naive algorithm which consists of an exhaustive enumeration of all
the partitions is then as efficient as any elaborated algorithm
when $N$ becomes large \cite{Mertens2}. Subsequently, even though the dynamics
studied in this paper seems a priori to be inappropriate to
the optimization problem, all the underlying mechanisms
responsible for out-of-equilibrium processes, especially aging
phenomena, come from the NP-hard nature of the problem.

So below $T_g$ any dynamical rules (satisfying detailed balance) lead to an aging regime before reaching the ground state.
In the following, we use K-spin-flip Metropolis rules ($1 \le K \le N$) defined as follows.
At each Monte-Carlo time step, a new configuration $\{ s_i' \}$ that differs from the current one $\{ s_i \}$ by at most $K$ spins (and at least one) is chosen randomly.
This new configuration is then accepted with a probability equal to the
Metropolis acceptance rates at temperature $T$:
\begin{equation}
W(\{ s_i \} \rightarrow \{ s_i'\})
= \left\{
\begin{array}{c}e^{- (E'-E)/T} \quad \mbox{if } E' > E
 \\ 1  \qquad \qquad \quad \; \mbox{if } E' \le E \\ \end{array}\right.
\label{Metr}
\end{equation}
Monte-Carlo time steps are separated by a physical time interval $\tau_{mc}=K/N$ in the natural time units of the system.
This ensures that each spin keeps a probability of the order of $1$ to be chosen within a unit time interval, even in the thermodynamic limit.

We show in the following that this model leads to a rich dynamic phase diagram, the control
parameter being the fraction $K/N$ of spins allowed to flip simultaneously.
This phase diagram can be discussed in light of both
the BMM and BTM studies at finite temperature.

\section{Simple realizations of trap models}

In this section, we study two different dynamical rules: a single-spin-flip dynamics ($K=1$) and a global dynamics involving full rearrangement ($K=N$).
Interestingly, these two limiting cases appear to be microscopic realizations of trap models, the former with an activated behavior and the latter with an entropic one.

\subsection{Single-spin-flip dynamics: activated traps}

It has been shown recently \cite{NPPtrap} that a single spin flip
dynamics naturally leads to an activated trap behavior {\it
\`a la} Bouchaud, in which the aging phenomenon comes from the
divergence of the average trapping time. Given the
density of state (\ref{energie}) with an exponential tail,
two dynamical ingredients are
responsible for such a behavior: on the one hand, the existence
of an horizon level below which the system has no choice but to
reemerge above it so as to continue its evolution; on the other
hand, instantaneous jumps into a randomly chosen new trap after
reemerging at the
horizon level, associated with a full decorrelation.

\begin{figure}[t]
\centering\includegraphics[width=8cm,clip]{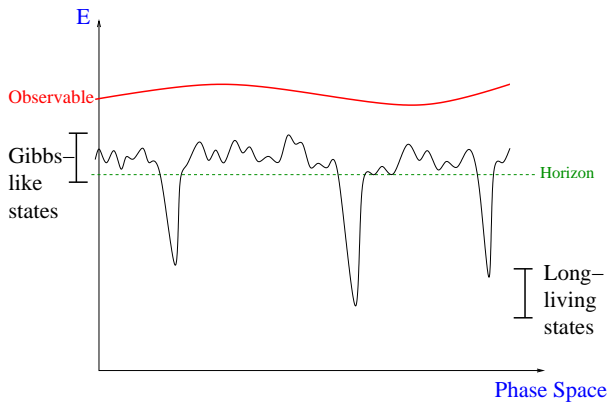}
\caption{Schematic representation of the phase space structure of the NPP model with single-spin-flip dynamics. The horizon level separates surface states and low energy states. An observable is smooth if it varies slowly between neighboring states.}
\label{smooth}
\end{figure}

The former appears naturally since when the energy is lower than:
\begin{equation}
E_h= T_g \ln(a_{\min}) \approx - T_g \ln N \label{horizon}
\end{equation}
with $a_{\min} \equiv \min (a_1,...,a_N)=O(1/N)$, a single spin
flip necessarily leads to a state whose energy is greater than
$E_h$. The latter is due to the combination of two properties.
First, low energy states are totally uncorrelated; second, the
travel time around $E_h$
between low energy states becomes negligible as time increases,
that is as lower and lower energy states are visited.

Interestingly, the need for reorganization around high energy
levels in order to go from one low energy state to another is
responsible for an equilibrium-like linear FDR with slope
$1/T$ for smooth observables, i.e.~observables like the magnetization
whose relative variation is of the order of $1/N$ after one spin flip
--see Fig.~\ref{smooth} for a schematic view. This law is observed even in the aging regime, $T$ being the temperature of the thermal bath.
This comes from the fact that the evolution of smooth observables is dominated by the sojourns among high energy states, where they can (almost) equilibrate.
Once in a deep state, these observables become frozen, but their typical value is indeed that given by Gibbs distribution, since it is determined by the high energy states visited just before falling into the trap.

\subsection{$N$-spin-flip dynamics and BMM behavior}

Let us consider now a global dynamics (i.e.~$K=N$)
such that all spins are flipped randomly (and simultaneously) at each step in order to find a new configuration.
The transition is then accepted or rejected
according to the Metropolis rates given in Eq.~(\ref{Metr}). As a
result, any configuration is a priori accessible from any other
(apart from the energetic constraint), which means that the horizon
level disappears, and the new configuration is in general completely
decorrelated from the old one. As moreover energies are
distributed exponentially, one can expect this model to be a microscopic
realization of the BMM. In the following, we propose more
quantitative arguments as well as numerical simulations to support
this statement.

\begin{figure}[t]
\centering\includegraphics[width=8cm,clip]{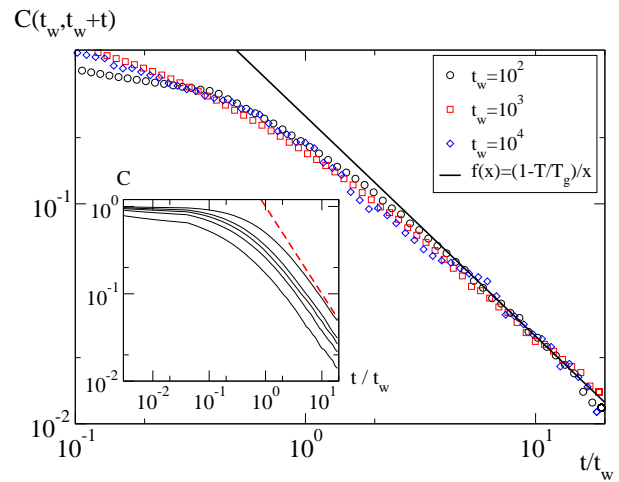}
\caption{Aging behavior of the correlation function $C(t_w,t_w+t)$ in the NPP with $K=N=50$, for $T=0.75 T_g$ and different $t_w$; data rescale as a function of $t/t_w$.
The full line is the analytical prediction Eq.~(\protect{\ref{eq-BMMlongt}}) for $t \gg t_w$.
Inset: correlation for different temperatures; from left to right: $T/T_g = 0.75,0.6,0.5,0.4,0$ ($t_w=10^3$). The dashed line is proportional to $t^{-1}$.}
\label{BMMlong}
\end{figure}

In all the numerical simulations, we have dealt with 
the autocorrelation function $C(t_w,t_w+t)$ between
time $t_w$ and $t_w+t$ defined by the average over the thermal histories
of the history-dependent correlation function
$C_{Single}(t_w,t_w+t)$:
\be
C(t_w,t_w+t) = \langle C_{Single}(t_w,t_w+t) \rangle
\ee
with
\begin{equation}
C_{Single}(t_w,t_w+t)=\frac{1}{N}\sum_{i=1}^{N} s_i(t_w) s_i(t_w+t)
\label{cor}
\end{equation}
This choice of correlation is usual in spin models, although other choices like the autocorrelation of the magnetization would be possible.
Actually, $C(t_w,t_w+t)$ is the autocorrelation of the observable $\sum_i \xi_i s_i$, where $\xi_i=\pm 1$ are quenched random variables.
We show in the appendix that the specific choice of the observable does not influence the main properties of the model, as long as the observable is smooth.

In the case of a full redistribution of the spins, this
autocorrelation reduces to the hopping correlation function $C^H(t_w,t_w+t)$, defined by the following history-dependent function:
\begin{equation}
C_{Single}^H(t_w,t_w+t)= \left\{
\begin{array}{cc}
1 & \quad {\rm if} \quad s_i(t_w+t)=s_i(t_w) \quad \forall i \\
0  & \qquad {\rm otherwise}
\end{array} \right.
\end{equation}
which precisely leads to the same correlation as in trap
models. The aging regime is characterized by the fact that the
correlation function $C(t_w,t_w+t)$ becomes a function
$\mathcal{C}(t/t_w)$ of the ratio $t/t_w$ only. For trap models
with exponential energy distributions, the asymptotic behavior of
the function $\mathcal{C}(t/t_w)$ for $t \ll t_w$ and for $t \gg
t_w$ is characteristic of the nature (entropic or activated) of
the dynamics \cite{Bertin}. In the NPP, numerical data show that (full) aging
is observed for any temperature $T<T_g$ as expected (see Fig.~\ref{BMMlong} for $T=0.75 T_g$).

\begin{figure}[t]
\centering\includegraphics[width=8cm,clip]{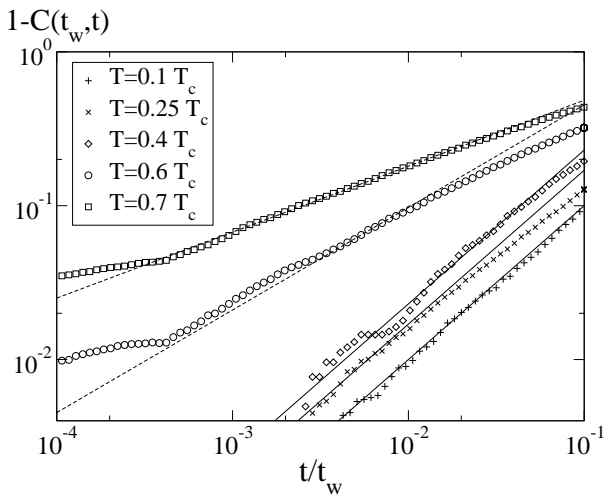}
\caption{Short time behavior of the correlation function $C(t_w,t_w+t)$ and onset of a singularity with exponent $(1-\mu)/\mu$ (full lines) for $\frac{1}{2} < \mu < 1$ ($\mu=T/T_g$), as given by
Eq.~(\protect{\ref{eq-short}})}.
\label{BMMcourt}
\end{figure}

In the long time limit ($t \gg t_w$), the asymptotic behavior
characteristic of the BMM, with a temperature independent
exponent, is recovered (Fig.~\ref{BMMlong}):
\begin{equation}
C(t_w,t_w+t) \sim \frac{t_w}{t}
\label{eq-BMMlongt}
\end{equation}

Actually, one can be more specific and compute the exact
asymptotic expression of the correlation function in the case of
Metropolis rates. One finds
$C(t_w,t_w+t) \approx (1-\mu)\, t_w/t$, where $\mu=T/T_g$ is the reduced temperature. This prediction fits well the
numerical data, as shown on Fig.~\ref{BMMlong}.

In the short time limit ($t \ll t_w$), an asymptotic analysis of
the correlation function in the BMM shows the onset of a
singularity for temperatures in the range $T_g/2<T<T_g$:
\begin{equation}
1-C(t_w,t_w+t) \sim \left\{
\begin{array}{ll}
\left(  \frac{t}{t_w} \right)^{(1-\mu)/\mu} \quad \quad \frac{1}{2} < \mu < 1\\
\\
\; \; \frac{t}{t_w} \qquad \qquad \qquad \mu < \frac{1}{2}
\end{array} \right.
\label{eq-short}
\end{equation}
More precisely, the singularity concerns the scaling function
$\mathcal{C}(u)$ in the limit $u \to 0$. This singularity clearly
appears also in the NPP with the $N$-spin flip dynamics, as shown
on Fig.~\ref{BMMcourt}.
This is the signature of an entropic-to-activated transition at
temperature $T_g/2$ \cite{Bertin}. Discrepancies at very
short time come from an exploration of states that are not
exponentially distributed due to finite size effects, but also from
finite time effects since correlation
functions are calculated analytically in the limit of asymptotically large times.

Different kinds of transitions between entropic and activated dynamics have been found in the context of the BMM, or of modified versions of this model \cite{Bertin}.
One, already mentioned in the introduction, is a crossover from an entropic to an activated regime as the system ages.
The heat bath temperature is kept constant in this process, and a characteristic time scale is associated to the crossover.
A second type of entropic to activated transition can be found also when varying temperature.
In the BMM with exponential energy density, such a transition appears when temperature crosses the value $T_g/2$; below $T_g/2$,
the dynamics is essentially dominated by entropic effects, while above this value (with still $T<T_g$ to remain in the aging regime) activated effects come at play.
This is seen in particular from the short-time behavior ($t \ll t_w$) of the aging correlation function which becomes singular above $T_g/2$, with an exponent $(1-\mu)/\mu$ reminiscent of (although different from) the exponent $1-\mu$ of the BTM \cite{Bertin}.
This transition is also present in the NPP with $K=N$, as seen from Eq.~(\ref{eq-short}), confirming the mapping between both models.

\section{Intermediate dynamics: $1 \ll K \ll N$}

We have seen in the previous section that the limiting dynamical rules ($K=1$ and $K=N$) correspond to the two simple kinds of trap models, namely BTM (activated) and BMM (entropic).
Since the ratio $K/N$ governing the dynamical rules can be varied (almost) continuously from $0$ to $1$, a crossover between both kinds of behaviour should be found.
Yet, this crossover is rather non trivial, as activated and entropic dynamics are qualitatively different.
In particular, one can expect the horizon level to play a major role in this change of dynamics.
The way the observables decorrelate may also lead to significant differences with respect to the standard trap picture.

\subsection{Differences with the previous rules}

\subsubsection{Slow decorrelation of smooth observables}

In simple trap models, one usually assumes that the value of the observable after a transition (a `jump') is completely decorrelated from the value it had before the jump.
This simple assumption, which might seem unrealistic at first sight, is indeed fulfilled by the single-spin-flip and the N-spin-flip dynamics, but for different reasons.
If $K=N$, it is clear that at each step, the new configuration is independent of the old one, so that observables immediately decorrelate.
For the case $K=1$, the mechanism appears to be more subtle: each time a new configuration is chosen, smooth observables typically decorrelate by a factor $(1-1/N)$.
Dynamics is dominated by low energy states (or traps), which are below the horizon level.
Once in such a trap, a single spin flip leads to a high energy state, and many subsequent flips are necessary in order to find a new trap.
Yet, the typical time spent wandering among these high energy states remains negligible compared to the time spent within traps.
So in terms of the effective dynamics between traps, the observables indeed fully decorrelate at each jump.

So what happens for $1 \ll K \ll N$?
In this case, two subsequent configurations are still highly correlated, and it is not clear either whether some relevant coarse-grained description could lead to an effective decorrelation.
One thus expects to observe some non trivial behaviors which may differ significantly from that of usual trap models.

\subsubsection{Influence of the horizon level}

The single-spin-flip dynamics studied previously has emphasized the fundamental role played by the horizon level.
On the other hand, this threshold completely disappears for $K=N$.
In the intermediate case $1 \ll K \ll N$, one can still define an horizon level, but this level is expected to drift towards lower energies as $K$ increases.
Using the same argument as above to determine the horizon --see Eq.~(\ref{horizon})-- one gets the following threshold energy:
\begin{equation}
E_h^K=-KT_g \ln N \qquad (K \ll N)
\label{horizonK}
\end{equation}
Below this level, the evolution is always activated: an energy barrier at least equal to $(E_h^K-E)$ has to be overcome when starting from an energy $E<E_h^K$.

\begin{figure}[t]
\centering\includegraphics[width=8cm,clip]{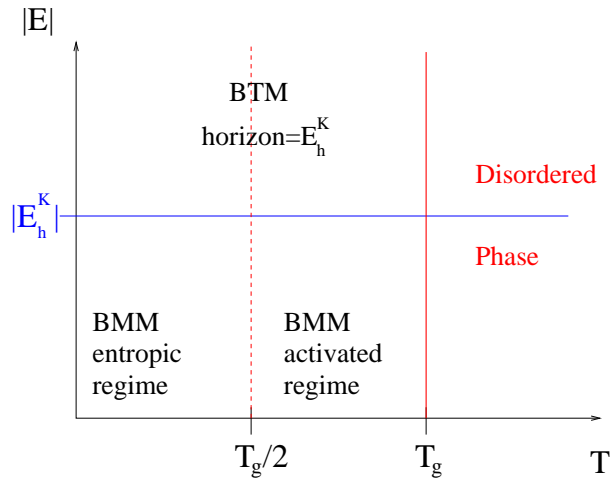}
\caption{Schematic view of the different dynamical regimes appearing in the NPP depending on $T$ and on the typical energy $E$ visited by the system (large values of $|E|$ correspond to deep energies and long times).}
\label{diagram}
\end{figure}

On the contrary, as long as the system visits states with energy $E$ well above $E_h^K$, the influence of the threshold should not be felt.
So one can guess that two different dynamical behaviors for temperatures above and below $T_g/2$
should still exist, as found also in the modified version of the BMM including a threshold \cite{Bertin}.
As a result, one expects to find schematically the three following regimes (see also Fig.~\ref{diagram}):

\begin{enumerate}

\item $E>E_h^K$, $T<T_g/2$.
This case resembles the entropic regime of the BMM, the only difference being that the magnetization decorrelates slowly, typically by a factor $(1-K/N)$ after each jump, at variance with the full decorrelation usually assumed in the BMM.

\item $E>E_h^K$, $T>T_g/2$.
As in case (1), the
dynamics is similar to that of the BMM; in this temperature range, activated effects become important, and observables also decorrelate slowly.

\item $E<E_h^K$. Once the horizon level is reached, the system has no choice
but to reemerge above it, which leads to an activated dynamics similar to that of the BTM.
No specific role is played by the temperature $T_g/2$ below the horizon, this activated regime is qualitatively the same in the whole range $T<T_g$.

\end{enumerate}

Thus, the NPP with a finite number of spin flips (i.e.~$K \ll N$) exhibits a crossover from an entropic regime to a activated one as time elapses, for any given temperature $T<T_g$.
Although it would have been interesting to investigate the properties
of the model within (and beyond) this crossover regime, we do not study them in the present paper, since the crossover time $t_{\times}$ is exponential in $K$ (more specifically, $t_{\times} \sim \exp(-E_h^K/T_g) \sim N^K$) which leads to time-consuming numerical simulations. We thus
focus on cases (1) and (2), corresponding to energies well above the horizon level. Subsequently, all the results reported
below correspond to times much smaller than the crossover time $t_{\times}$.

\subsection{Entropic versus activated dynamics}

\subsubsection{Qualitative approach}

\begin{figure}[t]
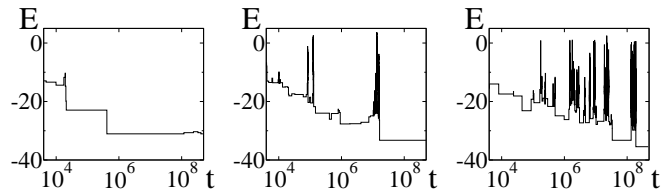

\includegraphics[height=2.4cm,clip]{singlerun35.eps}
\hfill
\includegraphics[height=2.4cm,clip]{singlerun50.eps}
\hfill
\includegraphics[height=2.4cm,clip]{singlerun65.eps}
\caption{Energy as a function a time for a single thermal
history at three different temperatures:
from left to right, $T/T_g=0.35$, $0.5$ and $0.65$ ($N=200$ and $K=5$). Returns to high
energy levels appear only for $T>T_g/2$. At this temperature, 
the dynamics is rather similar to the BTM, with in particular an apparent reversibility when plotted on a linear time scale \protect{\cite{NPPtrap}}.} 
\label{fig-Energie}
\end{figure}

Before giving quantitative arguments about correlation functions, let us propose a more intuitive understanding of the difference between entropic and activated dynamics.
To this aim, it is of
interest to plot the energy as a function of time for a single
thermal history. This is done on Fig.~\ref{fig-Energie} for three
different temperatures, $T/T_g=0.35$, $0.5$ and $0.65$. For
$T<T_g/2$, the energy decreases essentially in a monotonic way,
and the evolution is close to the zero temperature one: the dynamics remains entropic. On the
contrary for $T>T_g/2$, the system comes back many times to high
energy levels, as if it had to reemerge from a deep trap: activated events dominate the evolution, which is then rather similar to that of the activated trap model.

Note that qualitatively similar trajectories in energy space can be found in the case $K=N$ discussed in the previous section as well as in the BMM.
In this context, it has been proposed \cite{Bertin} to characterize
the type of dynamics by computing the
average energy $\langle E' \rangle_E$ reached in a transition between two different microscopic states, starting from a given
energy $E$:
\begin{equation}
\langle E' \rangle_E =\frac{\int_{- \infty}^0 dE' \; E' \, W(E
\rightarrow E')}{\int_{- \infty}^0 dE' \; W(E \rightarrow
E')}
\end{equation}
Using the Metropolis rules and assuming that the visited energies are still well above the horizon level, this quantity reads in the large $|E|$ limit:
\begin{equation}
\langle E' \rangle_E = E + \frac{2\mu-1}{1-\mu}\, T_g
\label{Efter}
\end{equation}
Below $T_g/2$, the energy is
lowered on average at each step, leading to an irreversible drift towards low energies.

On the contrary, above $T_g/2$ the energy is raised on average, i.e.~$\langle E' \rangle_E > E$, as long as $|E|$ is large enough ($E<0$).
So the energy variable performs a random walk in energy space, with a bias towards high energies, and --roughly speaking-- a reflecting boundary condition in $E=0$.
If time was counted in number of jumps, the random walk would then reach a steady state: energies tend to remain close to the boundary $E=0$, as seen on Fig.~\ref{fig-Energie}.
However, the larger $|E|$, the larger the sojourn time, so that large fluctuations away from the boundary (i.e.~at large $|E|$) dominate the real time dynamics.
So the probability to have energy $E$ at time $t$ does not reach any steady state, and drifts continuously towards low energies.
This competition between the bias towards high energy and the large trapping time of low energy leads to the onset of a singularity in the correlation function as discussed above.

Since a full analytical solution of the correlation function in the present model with $K \ll N$ seems difficult to reach, simple scaling arguments can be helpful in order to interpret numerical simulations.
Data shown on Fig.~\ref{fig-Energie} suggest that the type of scaling argument (or in other words, the relevant approximations) may be different for entropic and activated dynamics.
In the entropic range $T<T_g/2$, the instantaneous energy $E(t)$ can be decomposed into an average value (with deterministic evolution) and a fluctuating term with zero mean and a finite amplitude.
Even though fluctuations are not necessarily small, they do not dominate the dynamics and may be considered as a perturbation over the average deterministic evolution.
So it may be reasonable to think that a zeroth order approximation which would neglect fluctuations could yield some relevant results, in particular concerning the scaling behavior.

On the contrary, the dynamics for $T>T_g/2$ appears to be dominated by activated events during which the system visits high energy states.
Fluctuations are now driving the evolution, and cannot be considered anymore as a perturbation which could be ignored in a first step.
Indeed, since the system goes back frequently to superficial states at energy $E \approx 0$, the amplitude of the fluctuations (with respect to the average energy at time $t$) diverges with time.
So scaling arguments involving only average values cannot be used anymore.

\subsubsection{Entropic temperature range $T<T_g/2$} \label{sect-entrop}

Let us first
consider the case of zero temperature and estimate the aging
law for magnetization through a simple scaling argument.
Given that the magnetization decorrelates typically by a factor $(1-K/N)$ at each transition, one can compute the correlation $C_R$ after $R$ jumps.
Assuming $K \ll N$, we have in the large $R$ limit:
\begin{equation}
C_R \approx e^{-R\,K/N}
\label{corexp}
\end{equation}
with the prescription $C_{R=0}=1$.
From Eq.~(\ref{Efter}), we know that after each jump, the energy
decreases of $T_g$ on average, so that after $R$ jumps the energy difference
between times $t_w$ and $t_w+t$ is given by:
\begin{equation}
E(t_w+t)-E(t_w) \approx -R \, T_g
\label{difE}
\end{equation}

\begin{figure}[t]
\centering\includegraphics[width=8cm,clip]{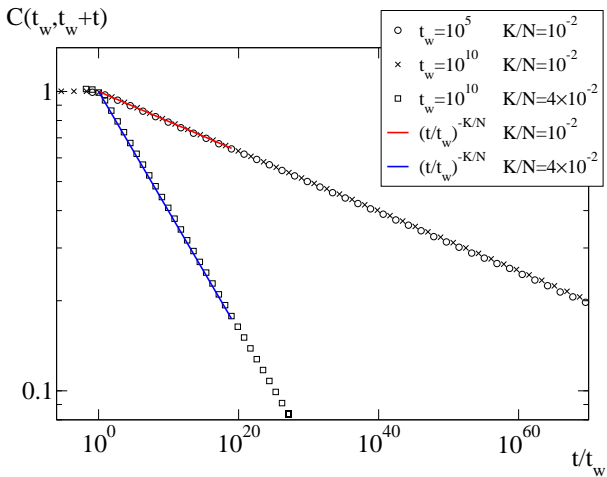}
\caption{Correlation function $C(t_w,t_w+t)$ up to very long time $t$ at zero temperature and for small values of the ratio $K/N$. Data were obtained from an efficient (event-driven) Monte-Carlo algorithm; the thermodynamic limit $N \to \infty$ is taken, so that the horizon level $E_h^K = -KT_g \ln N$ cannot be reached.}
\label{Maitresse}
\end{figure}

\begin{figure}[t]
\centering\includegraphics[width=8cm,clip]{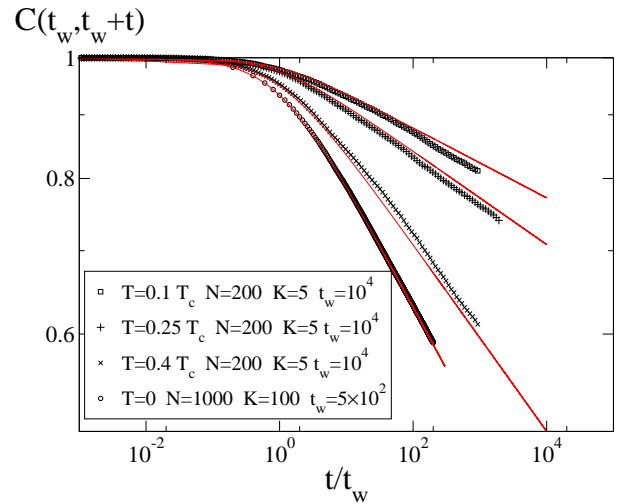}
\caption{Direct Monte-Carlo simulations of the correlation function in the range $T<T_g/2$, for small values of $K/N$. For $T>0$, $N=200$ and $t_w=10^4$; for $T=0$, $N=1000$ and $t_w=500$. It has been checked that typical energies remain well above the horizon level.
Lines are predictions given by Eq.~(\protect{\ref{Centre}}); no fitting parameter is used.}
\label{Corentro}
\end{figure}

From an energetic point of view, the NPP far above $E_h^K$ is expected to be
equivalent to the BMM. Subsequently, at an energy $E$ the corresponding 
trapping time is given by:
\be
\tau_E=\tau_{mc} \left( \int_{-\infty}^{E} \; dE' \rho(E') \right)^{-1}
\ee 
where $\rho(E')$ is given by Eq.~(\ref{energie}) and $\tau_{mc}=K/N$ --see Sect.~\ref{sect-opt}.
At low energy, this trapping
time reduces to:
\begin{equation}
\tau(E) \approx \tau_{mc}\, \frac{\beta_g e^{-E/T_g}}{\mathcal{N}}
\label{toto}
\end{equation}
where ${\cal N}$ is a factor coming from the distribution $\rho(E)$.
Since time $t_w+t$ is of the order of the typical trapping time of the
state currently visited, Eq.~(\ref{toto}) leads to the following relation between $t_w+t$
and $E$:
\begin{equation}
t_w+t \approx \frac{K}{N} \, \frac{\beta_g e^{-E/T_g}}{\mathcal{N}}
\end{equation}
This combined with Eqs.~(\ref{corexp}) and (\ref{difE}) gives:
\begin{equation}
C(t_w,t_w+t) \approx \left( \frac{t_w+t}{t_w} \right)^{-K/N}
\label{Centre0T}
\end{equation}
for the aging correlation function at zero temperature.
Although rather naive, this simple scaling argument is very well confirmed by long time simulations performed with an efficient event-driven algorithm, as seen on Fig.~\ref{Maitresse}.
The agreement with direct Monte-Carlo simulations
of NPP up to accessible times is also very satisfactory (see Fig.~\ref{Corentro}, lower curve).
Note that a modified version of the BMM which includes the property of slow decorrelation of the observable precisely leads to the same behavior as Eq.~(\ref{Centre0T}) at zero temperature \cite{Ip}.

For non zero but low enough temperature ($T<T_g/2$), so that activated processes do not dominate the dynamics, one can use exactly the same argument, simply modifying the time dependence of the average energy according to Eq.~(\ref{Efter}):
\begin{equation}
E(t_w+t)-E(t_w) \approx - \frac{1-2\mu}{1-\mu} \, R \, T_g \qquad \mu < \frac{1}{2}
\end{equation}
This gives:
\begin{equation}
C(t_w,t_w+t) \approx \left( \frac{t_w+t}{t_w} \right)^{-\eta \, K/N}
\label{Centre}
\end{equation}
with $\eta=(1-\mu)/(1-2\mu)$.
Here again, this simple estimation describes rather well the numerical simulations (Fig.~\ref{Corentro}).

So one sees that the law of decorrelation of the observable
between states, which is encoded in the ratio
$K/N$, has a dramatic impact onto the corresponding aging laws.
In particular, the specific behavior of the correlation function given by
Eqs.~(\ref{Centre0T}) and (\ref{Centre}) has important consequences
regarding the thermodynamic limit $N \to \infty$. According to the
dependence of $K$ on $N$, the correlation function is able or not
to decay at large times. Indeed, if $K \approx \alpha N$ with some positive
constant $\alpha$, $C(t_w,t_w+t)$ converges in the large $N$ limit
to a well defined scaling function which decays to $0$ for $t \to \infty$ --see Eq.~(\ref{Centre0T}).
On the contrary, if $K/N \to 0$ for $N \to \infty$ (say $K$ is
fixed), the system becomes unable to decorrelate in the
thermodynamic limit, and $C(t_w,t_w+t)$ remains equal to $1$.
Defining $\alpha$ as the limit when $N \to \infty$ of the ratio
$K/N$, and taking it as a control parameter, one sees that a
transition towards a state where ergodicity is broken
occurs at $\alpha=0$.

\subsubsection{Activated temperature range $T>T_g/2$}

As already mentioned in the introduction of this section,
the dynamics in the temperature range $T_g/2<T<T_g$ is
qualitatively different from that in the range $T<T_g/2$:
jumps to high energies, inducing large energy fluctuations,
play an important role in the evolution of the system.
In this case, a scaling argument based only on the deterministic
evolution of average values is not expected to be relevant.

Let us consider first the dynamics of the energy.
Since for $K \gg 1$ the horizon level $E_h^K$ is low,
a significant energy range above $E_h^K$ exists where
energy states are all mutually accessible, independent and
exponentially distributed.
Given the dynamical rules Eq.~(\ref{Metr}), the evolution of the energy
is precisely that of the BMM; for instance, one should find the same dynamic
probability distribution $P(E,t)$ --i.e.~the probability to have energy $E$ at time $t$.
Besides, once the threshold level is reached, a fully activated
dynamics typical of the BTM should be recovered.
So as far as the evolution of the energy is concerned, the situation is
very similar to the BMM with threshold studied in \cite{Bertin}.

\begin{figure}[t]
\centering\includegraphics[width=8cm,clip]{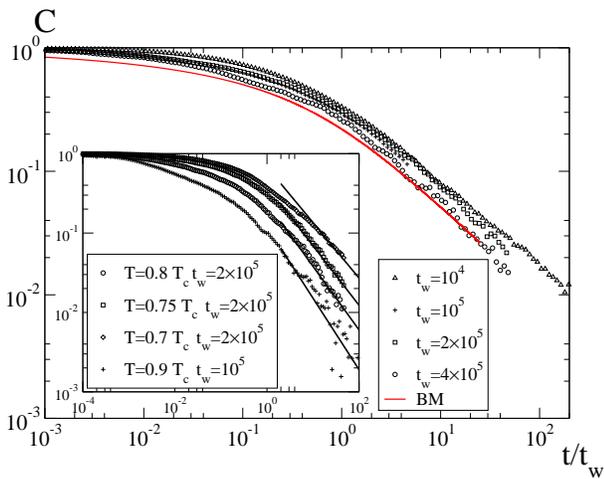}
\caption{Correlation function in the regime $T_g/2<T<T_g$ for the NPP with $1 \ll K \ll N$. Straight lines have slope $\mu=T/T_g$ on log-log scales, so as to compare with the activated behavior of the BTM.}
\label{BMMBM1}
\end{figure}

Turning to the correlation function of smooth observables, the situation is a bit more subtle.
Numerical data from the NPP model are shown on Fig.~\ref{BMMBM1}.
At variance with the usual results of the BMM, the long time tail ($t \gg t_w$) of the correlation function seems to behave as a power law with a temperature dependent exponent.
Yet, one must admit that these numerical data do not allow to draw a definitive conclusion on this point.

Since the magnetization decorrelates only by a factor $(1-K/N)$ (with $K \ll N$) at each transition, the evolution of the correlation should be that of the BMM with slow decorrelation.
Still, this model has not been studied in details in the literature, and
one could wish to know whether the behavior of the correlation function
resembles that of the hopping correlation function in one of the usual trap models.

\begin{figure}[t]
\centering\includegraphics[width=8cm,clip]{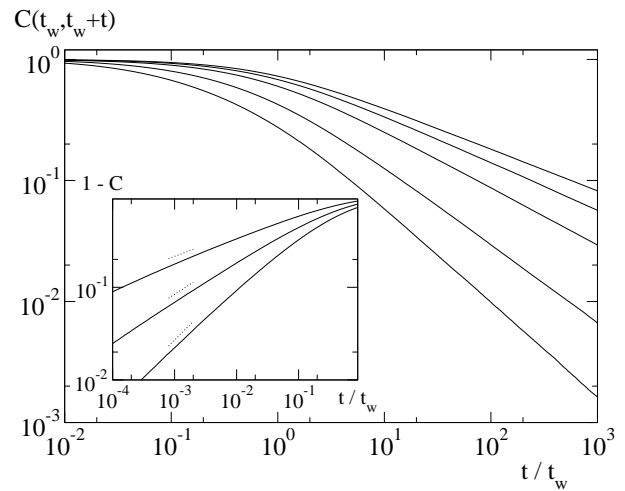}
\caption{Correlation function in the BMM with a decorrelation factor $(1-x)$; here, $x=0.1$.
From top to bottom: $T/T_g=0.52$, $0.55$, $0.6$, $0.7$ and $0.8$.
The long time tail becomes temperature dependent, with an exponent close to the value $-\mu$ found in the BTM, but deviations increase when $T \to T_g/2$.
Inset: short time behavior, with the typical exponent $(1-\mu)/\mu$ of the BMM (dashed).}
\label{BMMslowcor}
\end{figure}

To this aim, we have simulated directly a modified version of the BMM in which the observable is decorrelated only by a factor $(1-x)$ at each transition, assuming $x \ll 1$.
Numerical data are shown on Fig.~\ref{BMMslowcor}.
Interestingly, the behavior of the correlation function is numerically found to be reminiscent of both the BTM and the BMM.
The short time behavior ($t \ll t_w$) is very similar to that of the BMM:
a singularity $1-C \sim (t/t_w)^{\gamma}$ appears, with an exponent $\gamma$
very close to the value $(1-\mu)/\mu$ found for the hopping correlation
function of the BMM.
On the contrary, the long time tail becomes temperature dependent,
at variance with the usual BMM behavior given by Eq.~(\ref{eq-BMMlongt}), and in agreement with numerical
results found in the NPP --see Fig.~\ref{BMMBM1}.
The corresponding exponent is close to the exponent $-\mu$ typical of activated regimes, but significant discrepancies appear for $T$ close to $T_g/2$.
These discrepancies were somehow expected from a continuity argument,
since below $T_g/2$ the correlation function decays with a very small exponent $-\eta K/N$.

\section{Fluctuation-dissipation relations}

Generalizing the well-known equilibrium theorems, fluctuation-dissipation relations (FDR) 
have proven to be a very robust tool to study the out-of-equilibrium regime of glassy models \cite{Lejo}.
Given an observable $\mathcal{O}$, in most cases \cite{Lejo,Jole}, 
the two points correlation function between time $t_1$ and $t_2>t_1$,
\be
\mathcal{C}(t_1,t_2) \equiv \langle \mathcal{O}(t_1) \mathcal{O}(t_2) \rangle
\ee
and the response to a perturbation $h$ conjugate to  $\mathcal{O}$, 
\be
\mathcal{R}(t_1,t_2) \equiv \frac{\delta \langle \mathcal{O}(t_2) \rangle}{\delta h(t_1)}\Big\vert_{h=0}
\ee
are related through a FDR:
\be 
\mathcal{R}(t_1,t_2) = \frac{1}{T_{eff}(t_1,t_2)}\frac{\partial \mathcal{C}(t_1,t_2)}{\partial
t_1}
\ee
At equilibrium, $T_{eff}(t_1,t_2)$ is given by the thermal bath and all the system properties 
are time translational invariant.
Far from equilibrium, these FDR can be used to define a meaningful effective temperature \cite{CuKuPe,Jole,Rev}, 
as in the context of mean-field (or fully connected) spin-glass models \cite{Cuku}.
Since we are indeed dealing here with a fully connected spin model, it is then natural to study the FDR.
An important question to address is the temporal independence of $T_{eff}(t_1,t_2)$  
since only when $T_{eff}(t_1,t_2)$ does not depend on times can a unique and well-defined effective temperature be introduced.
In this case, introducing the integrated response,
\be
\mathcal{\chi}(t_1,t_2) \equiv \int_{t_1}^{t_2} dt' \, \mathcal{R}(t',t_2)
\ee
the FDR is said to be {\it linear} since the $t_2$-parametric plot $\mathcal{\chi}(t_1,t_2)$ vs. $\mathcal{C}(t_1,t_2)$ is a straight line
whose slope is given by $-1/T_{eff}$.

\subsection{FDR in the aging regime for $K=N$}

\begin{figure}[t]
\centering\includegraphics[width=8cm,clip]{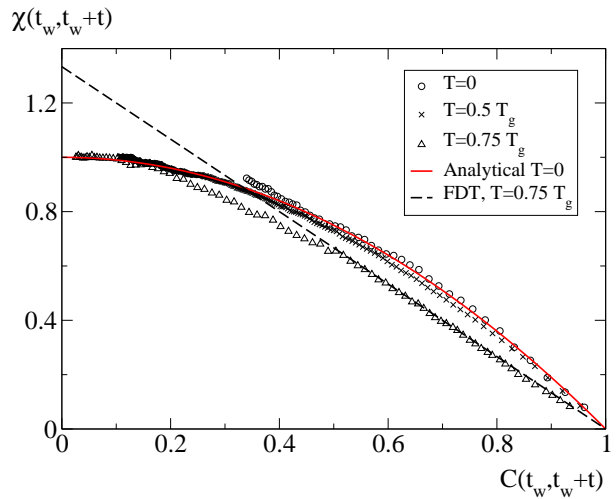}
\caption{Fluctuation-dissipation relations in the NPP for $K=N=50$ for different temperatures ($t_w=10^4$); $\chi$ is measured in units of $T_g$. Numerical data at $T=T_g/2$ remain very close to the zero temperature analytical relation of the BMM (full line). For $T>T_g/2$, the initial slope is given by $1/T$ (dashed line).}
\label{FDTsollich}
\end{figure}

The temperature $T_g/2$ separates the two different classes
of dynamics encountered in this model. It is then of interest to
compare the FDR in these two regimes. Numerical data concerning the FDR for the global dynamics ($K=N$) are shown on Fig.~\ref{FDTsollich}, in the case $T<T_g/2$.

As the NPP model with a global dynamics can be mapped
onto the BMM, one expects to recover the exact result found in the BMM at zero temperature \cite{Sollich} --see also
\cite{Ritort}--  which reads:
\begin{equation}
R(t_w,t_w+t)=-\frac{2}{T_g} \frac{\partial C(t_w,t_w+t)}{\partial
t}
\label{eq-FDR-BMM0T}
\end{equation}
$R(t_w+t,t_w)$ being the response associated to the autocorrelation
function $C(t_w,t_w+t)$.
In terms of the integrated response,
this FDR can be reformulated as:
\be
\chi(t_w,t_w+t) = \frac{1}{T_g} [1-C(t_w,t_w+t)^2]
\ee
taking into account the explicit expression of the correlation \cite{Sollich}.
Numerical data shown on Fig.~\ref{FDTsollich} are in good agreement with this analytical prediction.
Interestingly, this zero temperature solution
seems to be also valid for any temperature in the
`entropic range' $T<T_g/2$ (Fig.~\ref{FDTsollich}). As a result, below $T_g/2$, the FDR remains
{\it non linear}.

From a technical viewpoint, the autocorrelation (9) is associated to the observable $\sum_{i=1}^{N} \xi_i s_i(t)$, where $\{\xi_i\}$ is
a set of quenched independent random variables that can take the values
$\pm 1$. It is understood that all measured quantities are averaged over the realizations of $\{\xi_i\}$.
So the integrated
response $\chi(t_w,t_w+t)$ is numerically obtained in the NPP model by computing
\be
\chi(t_w,t_w+t) = \frac{1}{N\, h} \sum_{i=1}^{N} \xi_i s_i(t_w+t) \qquad t>0
\ee
where $h$ is a small external field that is switched on at time $t_w$.
This field $h$ is coupled to the spins via
a linear coupling term $V_h$ in the energy, involving also $\xi_i$:
\begin{equation}
E_h=E+ V_h \quad ; \qquad V_h= -h\sum_{i=1}^{N}\xi_i s_i
\label{VH}
\end{equation}

Above $T_g/2$, the FDR significantly depends on $T$ but seems to have
unchanged behavior in the two following limits: $t \ll t_w$
and $t \gg t_w$ (Fig.~\ref{FDTsollich}).
In the former, the slope of the curve $C$ vs.~$R$ is $T^{-1}$ (as in equilibrium) whereas in the latter, the slope apparently goes to $0$, which might be
interpreted as an infinite effective temperature.
Yet, such a temperature must be taken with care since the definition of effective temperatures from the local slope of non-linear FDR remains an unclear procedure. 

Apart from this question of effective temperature, it is quite interesting to see that one can discriminate between an entropic regime and an activated one in the BMM at finite temperature, from the initial slope of the FDR.
The entropic regime gives a T-independent slope
corresponding to the temperature $T_g/2$ that separates the two regimes,
whereas the activated regime gives a slope that corresponds to the
thermal bath temperature $T$.

\begin{figure}[t]
\centering\includegraphics[width=8cm,clip]{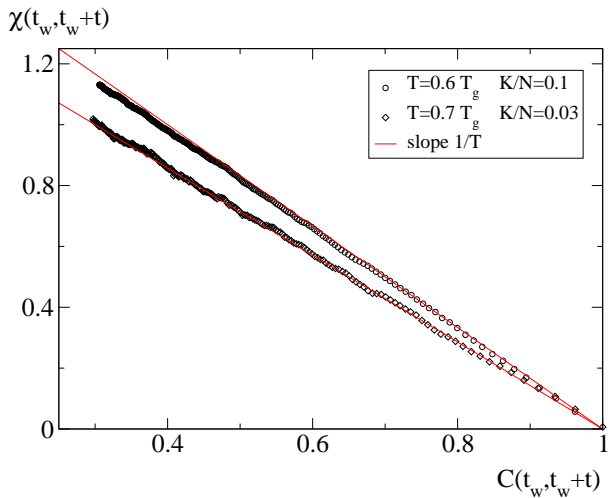}
\caption{FDR for $T>T_g/2$ in the aging regime of the NPP, for $1 \ll K \ll N$: a linear relation with slope $T^{-1}$ is observed.
Simulation parameters: $N=1000$ and $t_w=10^5$ for $T=0.6T_g$; $N=100$ and $t_w=2 \times 10^5$ for $T=0.7T_g$.}
\label{FDTact}
\end{figure}

\subsection{Out-of-equilibrium FDR for $K \ll N$}

Let us consider now the intermediate dynamics $1 \ll K \ll N$.
In the case $T>T_g/2$, one recovers when $K/N \rightarrow
0$ the same behavior as for $K=1$ \cite{NPPtrap}, namely a linear
FDR with a slope equal to $1/T$, where $T$ is the heat bath
temperature. Thus there is no violation of fluctuation-dissipation theorem, as seen on
Fig.~\ref{FDTact}. The mechanisms at play are essentially the same as for the activated dynamics found when
$K=1$ --see \cite{NPPtrap}-- so that this case need not necessarily be discussed in details.
Note that although simple trap models usually do not display linear FDR \cite{Sollich,Ritort,Sollich2}, such linear relations with slope $1/T$ as in equilibrium can indeed be found in trap models (at least for some particular observables) once a spatial structure is taken into account \cite{BertinFDT}.

\begin{figure}[t]
\centering\includegraphics[width=8cm,clip]{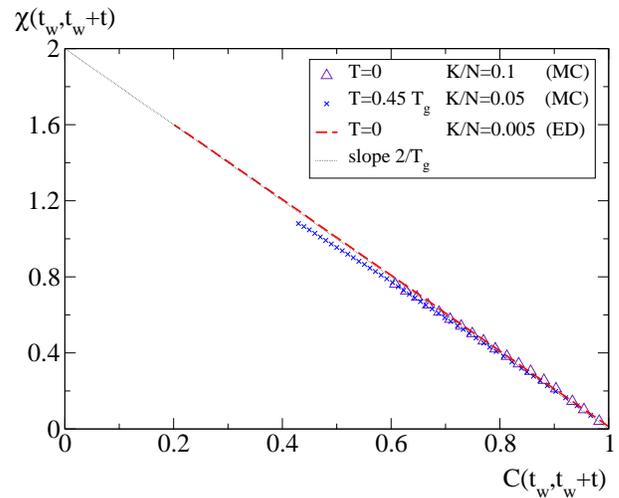}
\caption{FDR in the entropic temperature range $T<T_g/2$ in the NPP with $1 \ll K \ll N$.
Data were obtained by direct Monte-Carlo simulations (MC) or by the event-driven algorithm (ED); $t_w=10^3$ for MC and $10^{10}$ for ED.
Data converge to a linear FDR in the limit $K/N \to 0$ (and $t_w \to \infty$), with a slope $2/T_g$.
Thus in the regime, the effective temperature is equal to $T_g/2$.}
\label{FDTentro}
\end{figure}

On the contrary, the behavior of the system for $T<T_g/2$ is much
more surprising.
Indeed, as $K/N \rightarrow 0$, numerical
data on the zero temperature FDR converge to a linear relation
with slope $2/T_g$.
Results for different values of
$K/N$ and different temperatures below $T_g/2$ are gathered in Fig.~\ref{FDTentro}.
As mentioned above, the limit of linear FDR allows to define an effective temperature, which would be equal here to $T_g/2$ and
is thus temperature independent.
So it appears that this transition temperature between activated and entropic dynamics again plays an important role in the description of the low temperature phase.
It is also worth noticing that the initial slope (i.e.~for $C \approx 1$) of the non-linear FDR found for the global dynamics $K=N$ are the same as the slopes of the linear FDR in the case $1 \ll K \ll N$.

Here again, in this regime $T<T_g/2$, the NPP model presents strong similarities with the BMM modified to include slow decorrelation, as discussed in Sect.~\ref{sect-entrop}.
Indeed, it can be shown that this particular version of the BMM,
with a vanishing decorrelation of the observable at each transition,
also leads to a linear FDR with a slope $T_g/2$ \cite{Ip}

\subsection{Interpretation of the linear FDR}

Interestingly, this linear FDR with slope $2/T_g$ can be derived analytically in the whole regime $T<T_g/2$
for smooth observables.
The corresponding calculations are reported in Appendix.
In this section, we simply try to give a simplified picture of the 
physical mechanisms leading to this non-trivial effective temperature by considering 
the case of magnetization.

As seen in the case $K=1$ \cite{NPPtrap}, a key to the linearity of the FDR is the fact that the magnetization 
induced by the field $h$ is not influenced by the state in which the system was at time $t_w$, when the field was switched on.
In other words, the contribution to the magnetization of this initial state should be negligible.
This is indeed the case here, since the contribution is expected to be of the order of $Kh$, whereas the total 
magnetization should scale as $Nh$.
So in the limit $K/N \to 0$, the above contribution vanish. Notice that this property does not hold for
the BMM ($K=N$) since the initial state contribution is expected to be of the order of $Nh$. Subsequently, 
this case is not expected to give a linear FDT in agreement with the established results \cite{Sollich} (see
Eq.~(\ref{eq-FDR-BMM0T})).

Once this contribution is neglected, the linearity of the FDR can be given through a rather simple physical interpretation.
It simply means that the relaxation towards the non-zero magnetization induced by the field $h$ behaves in the same way as the relaxation towards zero starting from an arbitrary magnetization induced by the spontaneous fluctuations.
So the slope of the FDR is determined by the asymptotic value of the magnetization, a long time after the field was applied.
For $K=1$, the visits to superficial states induce asymptotically a constant magnetization $M^*$, since the a priori distribution:
\be
\rho(M) = \frac{1}{\sqrt{2\pi N}} \, e^{-M^2/2N}
\ee
is weighted by the Boltzmann factor $e^{hM/T}$; the average value of the resulting distribution is then $M^*=hN/T$.
On the other hand, the equal-time correlation (in the absence of field) $C(t,t)= \langle M^2 \rangle / N$ is equal to $1$, hence the slope $1/T$ of the FDR \cite{NPPtrap}.
One can expect this argument to be valid also in the regime $T>T_g/2$ for $1 \ll K \ll N$, thus accounting for the slope $1/T$ found on Fig.~\ref{FDTact}.

Turning to the case $T<T_g/2$, the asymptotic magnetization can be computed from the following argument.
Given the current state characterized by $E$ and $M$, one can compute the probability that the magnetization takes the value $M'$ after one transition, in the large $N$ limit --see Eqs.~(\ref{eq-PMEM}) and (\ref{eq-PMEM2}) in the Appendix.
This distribution is gaussian and independent of $E$; its average value $\overline{M}'$ can be identified with the most probable value:
\be
\overline{M}' = \left( 1-\frac{K}{N} \right) M + 2K \left( 1 - \frac{M^2}{N^2} \right) \, \frac{h}{T_g}
\ee
The asymptotic magnetization can be computed self-consistently by looking for the fixed point $\overline{M}'=M$.
Keeping only the first order in $h$, one can safely neglect the term $M^2/N^2$, since one expects $M \sim Nh$.
Solving the resulting equation and denoting the solution by $M^*$, one finds $M^* = 2Nh/T_g$.
So, as here again $C(t,t)=1$, the slope of the FDR is expected to be $2/T_g$.

So without entering into the details of the calculations developed
in Appendix, this simple argument already allows to get an intuitive
idea of the physical mechanisms responsible for the linearity of the FDR, and for the effective temperature $T_g/2$.

\section{Discussion}

\subsection{Effective temperature}

The effective temperature $T_g/2$ is very interesting for
several reasons.
On the one hand, it is surprising to see that this value is different from the usually
expected value $T_{\rm eff}=T_g$ which follows from mean-field
spin-glass models \cite{Monasson} and from Edwards-like arguments \cite{Nagel}. To be more specific, one may expect from these theories a value
\be
T_{\rm eff}^{-1} = \frac{\partial \ln \rho}{\partial E}
\ee
by associating $\ln \rho(E)$ with the complexity (or configurational entropy) of the system.
On the other
hand, similar results have been
found in a recent study of the Random Orthogonal Model (ROM)
\cite{Crisanti}, a spin-glass model with one-step replica symmetry
breaking solution. Although the results on the ROM were not
considered along the lines of the entropic-to-activated transition,
we believe that it would be of great interest to search for such a
transition in this model. To support this view, exponential tails proportional to $\exp(-Q/\lambda)$
in the distribution of heat exchanges $Q$ have been found in the ROM
\cite{Crisanti}. This suggests the existence of an underlying
exponential distribution of energy levels with a tail:
\be
\rho(E) \propto e^{E/\lambda} \qquad E \to -\infty
\ee
An interpretation based on a scenario of spontaneous and stimulated relaxation in glassy systems \cite{Ritortrel}, confirmed by numerical measurements of the FDR in the ROM \cite{Crisanti},
yields an effective temperature $T_{\rm eff}$ related to $\lambda$ through $T_{\rm eff}=\lambda/2$.
In other words, the relation between the effective temperature and the slope of the exponential distribution is the same as in the NPP with $1 \ll K \ll N$.
This suggests that the mechanisms at play in the NPP should be rather generic, on condition that energy levels are exponentially distributed.
However, this latter condition should not be so restrictive, since exponential distributions are expected for low lying energy states on the basis of extreme statistics \cite{BouchMez}.

\subsection{Influence of the energy distribution}

As already mentioned, two kinds of transitions between entropic and activated dynamics have to be distinguished in this model.
On the one hand, a transition can be found as a function of temperature when crossing the value $T_g/2$.
We have studied this transition in details within the NPP model.
On the other hand, a crossover between both dynamics also appears, at fixed temperature, when the energy of the system reaches the horizon level $E_h^K$.
Here the regime changes as function of time rather than temperature.
We did not study this phenomena in details within the NPP, as numerical simulations were difficult due to the large value of $|E_h^K|$.

The former transition is controlled by the ratio $\mu=T/T_g$ of the temperature $T$ and the characteristic energy $T_g$ of the exponential density of states $\rho(E)$.
On the contrary, the latter is due to the lack of direct paths between states lying below the horizon level: the system has to reemerge first above the horizon, hence the onset of activated dynamics.
So in the present model, different mechanisms are responsible for these two types of transitions.

Yet, the first mechanism, which is related to the shape of the energy distribution, could also lead to a crossover as a function of time, for a given temperature.
To this aim, one needs to consider a distribution $\rho(E)$ which is not purely exponential.
For instance, $\rho(E)$ could be composed of a first exponential part $\rho(E) \propto e^{E/T_g^{(1)}}$ for $E^*<E<0$, and a second one $\rho(E) \propto e^{E/T_g^{(2)}}$ for $E<E^*$.
Assuming $T_g^{(2)}/2 < T < T_g^{(1)}/2$, one then observes an entropic dynamics as long as $\overline{E}(t)$ remains well above $E^*$, and an activated one in the long time regime, when $\overline{E}(t)$ is below $E^*$.

Note that assuming on the contrary $T_g^{(1)}<T_g^{(2)}$, one finds a temperature range $T_g^{(1)}/2 < T < T_g^{(2)}/2$ such that the activated regime is found {\it before} the entropic one as the system ages (at constant temperature), which is rather counter-intuitive.
Clearly, this example using an energy density with two different exponential parts is a bit artificial, but it is interesting for pedagogical purposes.
A more natural example would be for instance a gaussian distribution:
\be
\rho(E) = \frac{1}{\sqrt{2\pi}} \, e^{-E^2/2E_0^2}
\ee
The steady-state distribution at temperature $T$, $P(E) \propto \rho(E)\, e^{-E/T}$, is also a gaussian shifted towards low energies, with average value $E_{st}=-E_0^2/T$ and the same variance $E_0^2$.
Assuming that we are in the low temperature regime $T \ll E_0$, one has $|E_{st}| \gg E_0$.
After a quench from high temperature to a low temperature $T$, the mean value $\overline{E}(t)$ is expected to drift slowly towards
the steady state value $E_{st}$ during the aging regime, while the variance remains essentially constant.
When $\overline{E}(t)$ becomes close to $E_{st}$ --say $\overline{E}(t) \approx E_{st}/2$-- one can linearize $\rho(E)$ around $\overline{E}(t)$, to find locally an exponential distribution:
\be
P(E,t) \propto e^{E/\lambda(t)}
\ee
with $\lambda(t)=E_0^2/|\overline{E}(t)|$.
This exponential approximation is valid as long as $|E-\overline{E}(t)| \ll E_0^2/T$.
In particular, if $|E-\overline{E}(t)| \approx T$, this approximation is fully justified.
It is then natural to define a reduced parameter $\mu(t)=T/\lambda(t)$, which plays the same role as $\mu$ in the BMM.
Since $|\overline{E}(t)|$ increases with time, so does $\mu(t)$; equilibrium is reached for $\mu=1$.
When $\mu(t)$ reaches the value $\frac{1}{2}$, one expects a crossover from an entropic dynamics to an activated one.

This behavior should hold more generally for any distribution $\rho(E)$ of the form:
\be
\rho(E) \sim e^{-\alpha |E|^{\gamma}} \quad E \to -\infty
\label{exp-gamma}
\ee
with $\gamma>1$.
On the contrary, distributions satisfying Eq.~(\ref{exp-gamma}) with $\gamma<1$ should lead to the reverse behavior, that is an activated regime followed by an entropic one, at least in some temperature range.
Note that such distributions with $\gamma<1$ exhibit a glassy behavior for any temperature \cite{BouchMez,Monthus}.

\section{Conclusion}

In this paper, we have established the dynamic phase diagram of the NPP model as a function of temperature and of the number $K$ of spins allowed to flip simultaneously.
The first result is that some particular dynamical rules lead to the behaviors found in both usual trap models: the case $K=1$ yields the activated trap model proposed by Bouchaud (BTM),
whereas the model with $K=N$ can be mapped onto the entropic version studied by Barrat and M\'ezard (BMM).
The former case has already been studied in \cite{NPPtrap}, but was recalled here for the sake of completeness.

In the intermediate range $1 \ll K \ll N$, the dynamics of the energy remains essentially the same as for $K=N$,
i.e.~of BMM type, within the time window accessible with numerical simulations. For longer time scales, a horizon level $E_h^K=-KT_g \ln N$ is expected to yield a crossover to an activated regime,
since states below $E_h^K$ behave as isolated traps.
Yet, the major difference with usual trap models is the presence of slow decorrelation of the observables: at each elementary transition between states,
the magnetization decorrelates typically by a factor $(1-K/N)$,
whereas full decorrelation in a single transition is assumed for usual trap models.
So the NPP can be mapped onto a modified version of the BMM which includes a slow decorrelation of the observable.

This extra property has important consequences.
On the one hand, in the entropic low temperature phase $T<T_g/2$, the correlation function decays only as $[(t_w+t)/t_w]^{-\eta K/N}$, with $\eta = (1-\mu)/(1-2\mu)$ and $\mu=T/T_g$, i.e.~much more slowly than the hopping correlation function.
Indeed, in the thermodynamic limit $N \to \infty$, an ergodicity breaking occurs (the correlation function does not decay to $0$) if $K$ is such that $K/N \to 0$.
Actually, for $N \to \infty$, the correlation decays to $0$ at large time only if $K \approx \alpha N$, with $\alpha >0$.
On the other hand, in the temperature range $T_g/2<T<T_g$, the short time behavior of the correlation is precisely that of the usual BMM, with the onset of a singularity above $T_g/2$,
but the long time (power law) tail becomes temperature dependent, with an exponent close to that of the BTM.
This result suggests that thermal activation may be in that case the only relevant mechanism to decorrelate the observables,
contrary to the activated phase of the BMM in which both thermal activation and entropic slowing down control the system.

In addition, we have studied the fluctuation-dissipation relation (FDR) in the NPP model.
In the limit $K \ll N$, the FDR becomes linear, and its slope depends on the temperature range considered.
For $T>T_g/2$, the slope is equal to $T^{-1}$, as in equilibrium and similarly to the case $K=1$.
On the contrary, for $T<T_g/2$, the slope is temperature independent and is equal to $2/T_g$.
Thus the effective temperature in this low temperature phase is $T_g/2$.

\subsection*{Acknowledgements}

We wish to thank J.~Kurchan, F.~Ritort and P.~Sollich for fruitful discussions, as well as 
M.~Droz for a careful reading of the manuscript. 
We are especially grateful to P.~Sollich for communicating to us his unpublished results. E.~B. has been supported by the Swiss National Science Foundation.

\section*{APPENDIX: Analytical developments on FDR}
In this appendix, we shall show, in the context of  a modified BMM,
how a well-defined effective temperature $T_g/2$ emerges for slow
decaying observables. To this end, we first study
the observable `magnetization'
($\mathcal{M}$) for a system microscopically composed of $N$ spins
and for which $K$ spins at most can be flipped. Subsequently,
magnetization is independent of the energy and the
energy evolution is the one given by the BMM endowed with
Metropolis dynamics (\ref{Metr}). We study the FDR for $T<T_g/2$,
taking the limits in the following order: $N \rightarrow \infty$,
$h \rightarrow 0$ and $K \rightarrow \infty$. Then, in the same
limits,  we generalize the results to the general case of smooth
observables that decorrelate by a factor $(1-K/N)$.

\subsection*{FDR for the magnetization}

\subsubsection*{Zero temperature case}

We begin by studying the problem when temperature
vanishes, choosing the magnetization as the (smooth) observable.
To derive the wished FDR, we need the relation between the magnetization $M$
before a jump and the magnetization $M'$ after it. The system is
assumed to be at an energy $E$, in the presence of an external field
$h$.

If $K$ spins are chosen to be flipped with probability
$\frac{1}{2}$, then $K/2$ spins are flipped on average.
Assuming that $K$ is large, fluctuations around this value $K/2$ can be neglected, and the effective dynamical rule is that $K/2$ spins chosen at random are flipped
(the new configuration found in this way is then accepted or rejected according to the Metropolis rate).
Given a magnetization $M$, the probability
for a spin to be in the up state is given by
\be
p_M=\frac{1}{2}+\frac{M}{2N}
\ee
The probability to flip a number $L \leq K/2$ of up spins reads:
\be
P_K(L)=\left(
\begin{array}{c}
  K/2 \\
  L
\end{array}
\right) p_M^L(1-p_M)^{\frac{K}{2}-L}
\ee
In the limit $1 \ll K \ll N$, using the gaussian limit of Bernoulli processes, one finds for the new
magnetization $M'$ at each Monte Carlo step the
following probability distribution:
\begin{equation}
\rho_K(M'|M)=\frac{1}{\sqrt{2 \pi
\alpha_M}} \, \exp \left( -\frac{(M'-(1-\frac{K}{N})M)^2}{2\alpha_M} \right)
\label{rhok}
\end{equation}
with $\alpha_M=8Kp_M(1-p_M)$. Taking into account the zero-temperature acceptance rate, the distribution of magnetization $P(M'|E,M)$ after
a transition is given by:
\bea
P(M'|E,M) &=& \frac{\rho_K(M'|M)}{\Gamma(E,M,h)} \times \\ \nonumber
&& \int dE' \, \rho(E') \, \Theta(E-hM-(E'-hM'))
\label{pmm}
\eea
where $\Theta(\cdot)$ is the Heavyside function, and $\Gamma(E,M,h)$ is the normalization factor that can be interpreted as
the escape rate from the state $\{E,M\}$:
\bea
\Gamma(E,M,h) &=& \int dM' \,\rho_K(M'|M) \times \\ \nonumber
&& \int dE'\, \rho(E') \, \Theta(E-hM-(E'-hM'))
\eea
One can then compute explicitly the distribution $P(M'|E,M)$:
\bea \label{eq-PMEM}
P(M'|E,M) &=& \frac{1}{\sqrt{2\pi\alpha_M}} \times \\ \nonumber
&& \exp \left(-\frac{(M'-(1-\frac{K}{N})M-\alpha_M
\beta_g h)^2}{2\alpha_M} \right)
\eea
which, interestingly, appears to be {\it independent} of the energy $E$.
This results in a decoupling between energy and magnetization.
Note that apart from the normalization factor,
the above calculation essentially amounts to multiply
the a priori distribution $\rho_K(M'|M)$ by a factor $\exp(\beta_g h (M'-M))$.
Given this distribution, we
can recursively compute the magnetization of the system after $R$
jumps. Assuming that the magnetization is $M_w$ at time $t_w$ when the field $h$ is applied,
and $M_R$ after $R$ jumps, one obtains the following
relations:
\begin{eqnarray}
\langle M_R M_w \rangle_0&=&a^R\langle M_w^2\rangle_0 \label{R1}\\
\langle M_R \rangle_h&=&a^R\langle M_w \rangle_0+2\beta_cNh(1-a^R) \label{R2}
\end{eqnarray}
with $a=1-K/N$. The subscript $0$ indicates that the
average is taken without any external field. Since
there is no spontaneous magnetization, one has $\langle M_w
\rangle_0 = 0$. In this case, and in this case {\it only},
one finds using also
$\langle M_w^2\rangle_0=N$:
\begin{equation}
\langle M_R \rangle_h=2\beta_g h (\langle M_w^2 \rangle_0-\langle
M_R M_w \rangle_0) \label{FDRR}
\end{equation}
Strictly speaking, this relation is not the FDR since the parametric variable
involved is the number $R$ of jumps. In order to convert Eq.~(\ref{FDRR})
into a relation involving times $t_w$ and $t_w+t$, we need to average it with
the distribution of time intervals $t$ given the number $R$ of jumps. Let us
consider this distribution in the presence of a small field $h$.
At leading order in $h$, it is the sum of the zero field
distribution and of terms proportional to $Kh$ (we do not write the exact distribution here for
the sake of simplicity). We then see that the limits have to be
taken in the following order: $N \to \infty$, $h \to0$,
$K \to \infty$. Thus, the average of Eq.~(\ref{FDRR}) with
that time distribution in the presence of $h$, combined with these
very limits, leads to:
\bea \nonumber
\frac{\partial \langle M(t_w+t) \rangle_h}{\partial
h} \Big\vert_{h=0} &=& \frac{2}{T_g} \left( \langle M(t_w+t)^2 \rangle_0 \right.\\
&-& \left. \langle M(t_w+t)M(t_w)\rangle_0 \right)
\eea
Making the following identifications:
\bea
\chi(t_w,t_w+t) &=& \frac{1}{N} \, \frac{\partial \langle M(t_w+t) \rangle_h}{\partial h} \Big\vert_{h=0}\\
C(t_w,t_w+t) &=& \frac{1}{N}\, \langle M(t_w+t)M(t_w)\rangle_0
\eea
we get the expected FDR, using also $C(t_w,t_w)=1$:
\be
\chi(t_w,t_w+t) = \frac{2}{T_g} \, [1-C(t_w,t_w+t)]
\label{eq-FDR0T}
\ee
Note that the order of the limits does not restrict so much the
validity domain of such a relation. Indeed, we have seen that $K$
does not need to be very large to get a long BMM regime.
And,
for finite $K$, the same relation as Eq.~(\ref{eq-FDR0T}) can be exactly  derived\cite{these}. 
The reason we have chosen
to present the large $K$ solution lies in the
simplicity of the relations between magnetization before and after a
jump. In this case, it should be noticed also that this FDR is only valid
for smooth observables with zero mean value at any time.
Such prescriptions onto the observables are very similar to
the ones needed to have a unique asymptotic FDR in the BTM \cite{Sollich2}.

\subsubsection*{Finite temperature}

We shall see how to extend Eq.~(\ref{eq-FDR0T}) to non-zero
temperatures. In this case, Eq.~(\ref{rhok}) is still
valid, whereas relation (\ref{pmm}) is modified due to the Metropolis rates. The distribution $P_T(M'|E,M)$ now reads:
\begin{eqnarray} \nonumber
&& P_T(M'|E,M) = \frac{1}{\Gamma^h_T(E,M)} \, \rho_K(M'|M) \times\\
&& \quad \left( \int_{-\infty}^{E-h(M-M')} \; dE' e^{\beta_g E'} \right. \\ \nonumber
&& \quad \left. +\int_{E-h(M-M')}^0 \; dE'
e^{-\beta(E'-hM'-(E-hM))} e^{\beta_g E'} \right)
\label{pmmT}
\end{eqnarray}
where $\Gamma^h_T(E,M)$ is the escape rate at temperature $T$ and with field $h$, defined by normalizing $P(M'|E,M)$.
Performing the
change of variables $x=E'-h(M'-M)$ in the integrals, one
gets:
\bea
P_T(M'|E,M) &=& \frac{\rho_K(M'|M)}{\Gamma^h_T(E,M)} \, e^{\beta_g h (M'-M)} \times \\ \nonumber
\left( \int_{-\infty}^E dx\, e^{\beta_g x} \right. &+& \left. \int_E^{h(M'-M)} dx\, e^{-\beta(x-E)+\beta_g x} \right)
\eea
As a result, the dependence on $h(M'-M)$ is essentially factorized, apart from the upper bound of the second integral.
In the regime $T>T_g/2$, this bound must play an important role, since the system visits quite often the high energy states.
On the contrary, in the opposite temperature range $T<T_g/2$, these superficial states are no longer visited in the long time regime, so that one can neglect the influence of this upper bound and safely replace it by $0$ since $h \ll |M'-M| \sim K$.
Within this approximation, the distribution $P_T(M'|E,M)$ is again computed by multiplying $\rho_K(M'|M)$ by the exponential factor $\exp(\beta_c h (M'-M))$, precisely as in the zero temperature case.

Thus, for {\it any} temperature lower than $T_g/2$,
the conditional distribution of magnetization after a transition is at large time the same as in the zero temperature case:
\begin{equation} \label{eq-PMEM2}
P_T(M'|E,M) = P(M'|E,M)
\end{equation}
Since moreover $P_T(M'|E,M)$ does not depend on $E$,
the derivation of the FDR is exactly the same as in
the zero temperature case. This shows that an effective
temperature equal to $T_g/2$ is expected in all the
entropic regime $T<T_g/2$.

\subsection*{Generalization to other smooth observables}

More generally, one can consider observables
$\mathcal{A}$ that decorrelate by a factor $(1-K/N)$ at each transition.
In this case, we
have the following relation between the value $A$ of the
observable before a jump and the value $A'$ after:
\be \langle A'A \rangle_0 = \left(1-\frac{K}{N}\right) \langle A^2 \rangle_0 
\label{obs} \ee
A natural stochastic evolution rule for such an observable is the following:
\be
A'=\left(1-\frac{K}{N}\right) A+ \zeta_A
\label{eq-A}
\ee
$\zeta_A$ is a gaussian white noise whose amplitude is
imposed both by the conservation of variance of $A$, that is
$\langle A'^2 \rangle=\langle A^2 \rangle$, and by the correlation given
in Eq.~(\ref{obs}).
Thus, at leading order in $K/N$, one can show that the variance of
$\zeta_A$ does not depend on the current value
$A$. The variance is given by:
\be
\langle \zeta_A^2 \rangle = \frac{2K}{N} \langle A^2 \rangle_0
\ee
Notice that we have implicitly chosen the case of observables that
are not correlated with the energy, since we assumed that $\zeta_A$ does
not depend on the energy.

Then, we can see that Eq.~(\ref{eq-A}) can be written as Eq.~(\ref{rhok}) with
$\alpha_M=2K \langle A^2 \rangle /N$. Thus, Eqs.~(\ref{R1}) and (\ref{R2})
become:
\bea
\langle A_R A_w \rangle_0&=&a^R\langle A_w^2\rangle_0 \\
\langle A_R \rangle_h&=&a^R\langle A_w \rangle_0+2\beta_c h \langle
A^2_w \rangle_0 (1-a^R) \eea
The end of the calculation is precisely the same as for the magnetization.

To conclude, this analysis shows that a generic smooth observable is expected
to satisfy a linear FDR with an effective temperature $T_g/2$ when
$T<T_g/2$. Moreover, we believe that the stochastic rule (\ref{eq-A}) is not an unavoidable
ingredient for the emergence of the effective temperature. Indeed,
as we already mentionned, il can be shown that magnetization verifies (\ref{eq-FDR0T})
even in the finite $K$ case for which the gaussian law (\ref{rhok}) does not apply.

\end{document}